\documentclass[9pt,twocolumn]{article}
\usepackage{fixltx2e}
\pdfoutput=1

\usepackage{extsizes,tabularx}

\newcolumntype{Y}{>{\small\center\arraybackslash}X}
\newcolumntype{W}{>{\small$}c<{$}} 
\newcolumntype{Z}{>{\small\centering\arraybackslash$}X<{$}} 

\usepackage[super,sort&compress,comma]{natbib} 
\usepackage[version=3]{mhchem}
\usepackage[left=1.5cm, right=1.5cm, top=1.785cm, bottom=2.0cm]{geometry}
\usepackage{balance}
\usepackage{widetext}
\usepackage{times,mathptmx}
\usepackage{sectsty}
\usepackage{graphicx} 
\usepackage{lastpage}
\usepackage[format=plain,justification=raggedright,singlelinecheck=false,font={stretch=1.125,small,sf},labelfont=bf,labelsep=space]{caption}
\usepackage{float}
\usepackage{fancyhdr}
\usepackage{fnpos}
\usepackage[english]{babel}
\usepackage{array}
\usepackage{droidsans}
\usepackage{charter}
\usepackage[T1]{fontenc}
\usepackage[usenames,dvipsnames]{xcolor}
\usepackage{setspace}
\usepackage[compact]{titlesec}
\usepackage{amsmath}
\usepackage{amssymb,amsfonts,amsthm,bm}

  \usepackage{xcolor}
    \usepackage{lipsum} 
\usepackage[draft]{todonotes}   

\usepackage{epstopdf}

\definecolor{cream}{RGB}{222,217,201}

\begin{document}

\thispagestyle{plain}

\makeFNbottom
\makeatletter
\renewcommand\LARGE{\@setfontsize\LARGE{15pt}{17}}
\renewcommand\Large{\@setfontsize\Large{12pt}{14}}
\renewcommand\large{\@setfontsize\large{10pt}{12}}
\renewcommand\footnotesize{\@setfontsize\footnotesize{7pt}{10}}
\makeatother

\renewcommand{\thefootnote}{\fnsymbol{footnote}}
\renewcommand\footnoterule{\vspace*{1pt}%
\color{cream}\hrule width 3.5in height 0.4pt \color{black}\vspace*{5pt}} 
\setcounter{secnumdepth}{5}

\makeatletter 
\renewcommand\@biblabel[1]{#1}            
\renewcommand\@makefntext[1]%
{\noindent\makebox[0pt][r]{\@thefnmark\,}#1}
\makeatother 
\renewcommand{\figurename}{\small{Fig.}~}
\sectionfont{\sffamily\Large}
\subsectionfont{\normalsize}
\subsubsectionfont{\bf}
\setstretch{1.125} 
\setlength{\skip\footins}{0.8cm}
\setlength{\footnotesep}{0.25cm}
\setlength{\jot}{10pt}
\titlespacing*{\section}{0pt}{4pt}{4pt}
\titlespacing*{\subsection}{0pt}{15pt}{1pt}

\fancyfoot{}
\fancyfoot[LO,RE]{\vspace{-7.1pt}\includegraphics[height=9pt]{head_foot/LF}}
\fancyfoot[CO]{\vspace{-7.1pt}\hspace{13.2cm}\includegraphics{head_foot/RF}}
\fancyfoot[CE]{\vspace{-7.2pt}\hspace{-14.2cm}\includegraphics{head_foot/RF}}
\fancyfoot[RO]{\footnotesize{\sffamily{1--\pageref{LastPage} ~\textbar  \hspace{2pt}\thepage}}}
\fancyfoot[LE]{\footnotesize{\sffamily{\thepage~\textbar\hspace{3.45cm} 1--\pageref{LastPage}}}}
\fancyhead{}
\renewcommand{\headrulewidth}{0pt} 
\renewcommand{\footrulewidth}{0pt}
\setlength{\arrayrulewidth}{1pt}
\setlength{\columnsep}{6.5mm}
\setlength\bibsep{1pt}

\makeatletter 
\newlength{\figrulesep} 
\setlength{\figrulesep}{0.5\textfloatsep} 

\newcommand{\topfigrule}{\vspace*{-1pt}%
\noindent{\color{cream}\rule[-\figrulesep]{\columnwidth}{1.5pt}} }

\newcommand{\botfigrule}{\vspace*{-2pt}%
\noindent{\color{cream}\rule[\figrulesep]{\columnwidth}{1.5pt}} }

\newcommand{\dblfigrule}{\vspace*{-1pt}%
\noindent{\color{cream}\rule[-\figrulesep]{\textwidth}{1.5pt}} }

\newcommand{\bnd}{\gamma}
\newcommand{\mbx}{\bf x}

\makeatother

\twocolumn[
  \begin{@twocolumnfalse}
\vspace{3cm}
\sffamily
\begin{tabular}{m{4.5cm} p{13.5cm} }
 & \noindent\LARGE{\textbf{ Dynamics of a multicomponent vesicle in shear flow $^\dag$}} \\
\vspace{0.3cm} & \vspace{0.3cm} \\

 & \noindent\large{Kai Liu\textit{$^{a}$}, Gary R. Marple\textit{$^{b}$}, Shuwang Li\textit{$^{c}$}, Shravan Veerapaneni\textit{$^{b}$} and John Lowengrub\textit{$^{a}$}} \\

 & \\
 & \noindent\normalsize{We study the fully nonlinear, nonlocal dynamics of two-dimensional multicomponent vesicles in a shear flow with matched viscosity of the inner and outer fluids.
 Using a nonstiff, pseudo-spectral boundary integral method,  we investigate dynamical patterns induced by inhomogeneous bending for a two phase system. Numerical results reveal that there exist  novel phase-treading and tumbling mechanisms that cannot be observed for a homogeneous vesicle. In particular, unlike the well-known steady tank-treading dynamics characterized by a fixed inclination angle, here the phase-treading mechanism leads to unsteady periodic dynamics with an oscillatory inclination angle. When the average phase concentration is around 1/2, we observe tumbling dynamics even for very low shear rate, and the excess length required for tumbling is significantly smaller than the value for the single phase case. We summarize our results in phase diagrams  in terms of  the excess length, shear rate, and concentration of the soft phase. These findings go beyond the well known dynamical regimes of a homogeneous vesicle and highlight the level of complexity of vesicle dynamics in a fluid due to  heterogeneous material properties.}


\end{tabular}

 \end{@twocolumnfalse} \vspace{0.6cm}

  ]

\renewcommand*\rmdefault{bch}\normalfont\upshape
\rmfamily
\section*{}
\vspace{-1cm}


\footnotetext{\textit{$^{a}$~Department of Mathematics, University of California in Irvine, Irvine, CA, U.S.  Tel: (949) 751-9700; E-mail:  lowengrb@math.uci.edu}}
\footnotetext{\textit{$^{b}$~Department of Mathematics, University of Michigan, Ann Arbor, MI, U.S. Tel: (734)-936-9963; E-mail: shravan@umich.edu }}
\footnotetext{\textit{$^{c}$~Department of Applied Mathematics, Illinois Institute  of Technology, Chicago, U.S.}}

\footnotetext{\dag~Electronic Supplementary Information (ESI) available.}




\section{Introduction}
As the principal components of living cells and organisms, membranes contain a mixture of materials such as lipids and cholesterol \cite{lip1,alb1}.   From a physical point of view, this inhomogeneous system may  go through a phase decomposition process to reach a lower energy state \cite{VeatchKeller} e.g., form coexisting phase domains (micro-domains or rafts) with distinct compositions. Experiments of Giant Unilamellar Vesicles (GUVs) show that membranes initially containing ternary mixtures of lipid components and cholesterol indeed separate into binary ordered ($L_o$) and disordered ($ L_\alpha$) liquid phases \cite{Baumgart2003,Baumgart2005}. The domain separations and structural rearrangements in membranes are also coupled with shape deformations and even topological changes \cite{Baumgart2003,Baumgart2005,Muk2004,McM2005}. 

When the mixed components are decomposed into phase domains, the mechanical responses of the vesicle (e.g. bending stiffness) may depend on the local concentration of these phase domains, i.e. budding and fission of a multiphase vesicle \cite{BenAmar2004,BenAmar2006,BenAma20062,Staneva2004}.  While there have been many theoretical and numerical studies on the bending energy of lipid bilayer membranes (e.g. see the reviews \cite{Nelson,lip2,lip1,Pozrikidis1995,Pozrikidis1992,Baumgart20142} and the references therein, and the recent papers \cite{Veerapaneni2016,Rahimian2010,Misbah2013, Misbah2016,Elliott2011,Salac2011}),  studies on inhomogeneous vesicles in fluids are more limited\cite{Baumgart2014,Aland2015b, Baumgart20152,Baumgart2015,Hora2016, Elliott2010,Tsutomu2011}.  Here, we investigate the dynamics of a multi-component vesicle in a two-dimensional shear flow.   In our model,  the energy of the system includes two parts: the bending energy with bending stiffness depending on the local concentration of surface phases \cite{Jinsun} and the line energy of the surface phases taking a Ginzburg-Landau form \cite{LipowskyR1992,Seifert1993,Julicher1993,Julicher1997,Elliott2010,Elliott2013,Fried2013}.  The stress jump across the membrane is naturally coupled with the phase decomposition process \cite{Jinsun,Elliott2010,Elliott2013,Elliott20102}. Such a continuum approach allows computation to reach large length and time scales than discrete approaches such as Monte Carlo methods \cite{Kumar2001,Kumar2005,Wallace2005}, dissipative particle dynamics \cite{Shillcock2005,NoguchiG2006,Jou1997,Grafmuller2007,SmithK2007,GaoL2008} or molecular dynamics \cite{HuangK2006,Markvoort2007}.


In this paper, using an integral equation method \cite{Jinsun}, we extensively explore dynamical patterns induced by inhomogeneous bending for a two phase system. We assume that the fluids inside and outside the vesicle have the same viscosity and vesicle membrane initially contains a uniform mixture of the two phases. In this work, we focus primarily on three parameters: excess length of the vesicle (defined in Sec.\ref{methods}), average concentration of the soft phase, and the applied shear rate. Our numerical results reveal that  phase distribution and inhomogeneous bending moduli lead to rich and novel dynamics even in simple shear flows. For a nearly circular vesicle, we observe that there exists a critical shear rate, above which the phase domain will start to move along the interface and the vesicle morphology will oscillate periodically. Unlike the well-known steady tank-treading dynamics characterized by a fixed inclination angle \cite{kraus1996fluid, de1997deformation}, it leads to unsteady periodic dynamics with oscillatory inclination angle. We call this phase-treading. Furthermore, the vesicle can move off the center position as the shape of the vesicle changes with the phase distribution. The critical shear rate depends linearly on the bending energy gap between the two phases. 

For a vesicle with large excess length, we observe tumbling dynamics in addition to the phase-treading and tank-treading dynamics. In particular, when the average phase concentration is around 1/2, tumbling of a vesicle can be observed even at low shear rates and the excess length required for tumbling is significantly smaller than that for a single phase case. This novel tumbling mechanism is mainly due to the inhomogeneous bending that the shape of the vesicle intends to bend inward at the soft phase region.   To the best of our knowledge,  no such tumbling mechnism has been reported so far.  We summarize our results in phase diagrams in the parameter plane (excess length, shear rate, and concentration of the soft phase). These findings go beyond the well-known dynamical regimes of a homogeneous vesicle and  highlight the level of complexity of vesicle dynamics in  fluids due to inhomogeneous bending.

\section{Materials and Methods}
\label{methods}
Consider a closed vesicle whose interior and exterior is filled with a fluid of viscosity $\eta$.  
Let $\bnd$ denote the boundary of the vesicle. We nondimensionalize the system using a characteristic length of the membrane $R$ (radius of an equivalent circular vesicle with the same area) and a characteristic time $\tau=\eta R^3/B$ where $B$ is a characteristic bending stiffness.  The excess length $\Delta=L/R-2\pi$, where $L$ is the total arc length.
 
\subsection{Flow field}  
We assume that the ambient fluid is governed by the Stokes equations,
\begin{eqnarray}
\nabla \cdot {\bf T} &=& 0  ~~~~\rm{and}~~~~\nabla\cdot{\bf u} = 0 \quad \mbox{in} \quad \mathbb{R}^2\setminus\bnd,
\label{eq:Stokes}
\end{eqnarray}
where ${\bf T}$ and ${\bf u}$ are the fluid stress and velocity, respectively. Across the interface $\bnd$, the velocity is continuous:
\begin{equation}
[\![{\bf u}]\!]_\bnd \; := \; {\bf u}|_{\bnd, \,\text{int}} -{\bf u}|_{\bnd, \,\text{ext}} = 0.
\label{eq:jump0}
\end{equation}
On the other hand, the hydrodynamic stress sustains a jump given by the Laplace-Young condition:
\begin{equation}
[\![{\bf T} {\bf n}]\!]_{\bnd}={\bf f},
\label{eq:jump1}
\end{equation}
where ${\bf f}$ is the total membrane force and  ${\bf n}$ is the outward normal to the interface $\bnd$.

In this work, we restrict our attention to simple shear flows, wherein, the far-field boundary condition is given by 
\begin{equation}
{\bf u}_\infty ({\bf x}) = S (x_2, 0)~~~\quad~~~{\rm  as} \quad |\!| \mbx |\!| \rightarrow \infty,
\label{far field}
\end{equation} 
for some point ${\bf x} = (x_1, x_2)$ in the fluid domain and   $S$ is the constant shear rate.  Finally, imposing a no-slip boundary condition at the vesicle boundary yields the following kinematic condition: 
\begin{equation}
\dot{\mbx} = {\bf u} \quad \text{on} \quad \bnd,
\label{interface equation}
\end{equation}
where $\dot{\mbx}$ is the velocity of a material point on the vesicle membrane.  The local inextensibility reads
\begin{equation}
\nabla_{{\bf s}}\cdot {\bf u} =0,
\label{inextensible}
\end{equation}
where ${\bf s}$ is the counter clockwise tangential of interface $\bnd$ .
\subsection{Material Field}  For simplicity, we focus our study on a vesicle whose membrane is composed of two phases only (e.g. lipid components). Let $\psi(s,t)$ denote the mass concentration of one phase, where  $s$ is the arclength parameterizing  the moving interface $\bnd$; the concentration of the other phase then is $1-\psi(s,t)$. We assume that no chemical reactions occur and the two phases are distributed only on the interface $\bnd$. Therefore, for each phase the total mass is conserved, 
\begin{equation}
\label{eq:concentration}
{M_\psi }(t) = \int\limits_{\bnd } {\psi (s,t) \, d\bnd(s) }  = {M_\psi }(0),
 \end{equation}
where we assume the surface density of each phase equals to one for simplicity. Thus, the evolution of $\psi$ is governed by a convection-reaction-diffusion equation. In Eulerian coordinates, the local form reads
\begin{equation}
\label{eq:flux}
{\psi _t} + {\bf{u}} \cdot \nabla \psi  - {\bf{n}} \cdot \nabla {\bf{u}} \cdot {\bf{n}}\, \psi  = {\nabla _{\bf  s}} \cdot {{\bf{J}}},
 \end{equation}
where  ${\bf{J}}$ is the surface flux derived below in eqn (\ref{flux
  constitutive form}), and operator $\displaystyle
 \nabla_{\bf  s}=({\bf I-nn})\nabla$. The expression
$-{\bf n}\cdot\nabla{\bf{u}}\cdot{\bf n}=\nabla_{\bf s}\cdot{\bf{u}}_{\bf  s}+H{\bf{u}}\cdot{\bf n}$,
where ${\bf{u}}_{\bf  s}=({\bf I-nn}){\bf u}$ is the tangential velocity on $\bnd$, and $H$ is the local mean curvature. For an incompressible velocity field, ${\bf n}\cdot\nabla{\bf{u}} \cdot{\bf n}$ describes the  local rate of change of the interfacial area (or the arclength in 2D). Correspondingly, this term in eqn (\ref{eq:flux}) describes the change in
$\psi$ due to interface stretching. Since we assume that the vesicle membrane is locally inextensible, equation (\ref{eq:flux}) reduces to a diffusion equation, which in Lagrangian coordinates is
\begin{equation}
\label{eq:phase}
\psi_t={\nabla _{\bf  s}} \cdot {{\bf{J}}}.
\end{equation}

\subsection{Constitutive Relations}  
  We consider the energy of the system 
\begin{eqnarray}
E &=& E_b + E_p + E_\psi \quad {\rm with}
\label{energy11}\\
 E_b&=&\displaystyle \frac{1}{2}\int_{\bnd}
  B(\psi)H^2~d\bnd, \label{Eb1}\\ 
E_p&=&\int_{\bnd} \Lambda ~d\bnd,
\label{Ep1}
\\
~~~{\rm and}~~~~E_\psi&=&\frac{a_0}{\epsilon}\int_{\bnd} \left(g(\psi)
+\frac{\epsilon^2}{2}|\nabla_s \psi|^2\right)~d\bnd, \label{Ef1}
\end{eqnarray}
where $E_b$ is the bending energy associated with inhomogeneous bending stiffness of the membrane $B(\psi)$ that depends on the local phase concentration $\psi$. $E_p$ is the energy due to membrane tension $\Lambda$, which can be viewed as a Lagrange multiplier that enforces the local inextensibility constraint. 

$E_\psi$ models the line energy associated with the surface phases. The function $g(\psi)$ takes the form of a double-well potential $\displaystyle g(\psi)=\frac{1}{4}\psi^2(1-\psi)^2$, with the two minima giving stable phases at $\psi=0$ and $\psi=1$.  $\epsilon$ is a small parameter (taken to be a constant for simplicity) that measures the excess energy due to surface gradients. It is chosen such that $E_\psi$ approaches a finite constant when $\epsilon\rightarrow 0$. The parameter $a_0$ characterizes the size of the line energy. 

The jump in the stress across the interface can be derived by taking variation of the total membrane energy with respect to the interface position, 
\begin{equation}
{\bf f}=-\frac{\delta E}{\delta \bnd},
\label{stressconstitutiveform}
\end{equation}
which is given by
\begin{eqnarray}
 \frac{\delta E  }{\delta \bnd } &=&  \left( - (B H)_{ss} - \frac{B}{2}H^3+\frac{a_0 }{\varepsilon}\left(g - \frac{{\varepsilon ^2 }}{2} \psi_s^2 \right)H + \Lambda H \right){\bf{n}}\nonumber \\ 
&~&~~~~+ \left(\left(\frac{a_0}{\varepsilon}\left( g' - \varepsilon^2  \psi_{ss} \right) +\frac{B'}{2}H^2 \right)\psi_s -\Lambda_s \right) {\bf s}
\label{energyVAR}
\end{eqnarray}
where $(\cdot)_s$ and $(\cdot)'$ indicate differentiation with respect to the arclength and $\psi$ respectively.

We define the surface flux as 
\begin{equation}
{\bf J}=\nu\nabla_{\bf s} \mu,
\label{flux constitutive form}
\end{equation}
where $\nu$ is a positive mobility coefficient and the chemical potential $\mu$ is defined by
\begin{equation}
\mu = \frac{\delta E}{\delta \psi} = \frac{1}{2} B^\prime  H^2 + \frac{a_0}{\epsilon}(g^\prime - \epsilon^2 \psi_{ss}). 
\end{equation}

In summary, the model requires us to solve the Stokes equations (\ref{eq:Stokes}) subject to  local inextensibility condition (\ref{inextensible}) and boundary conditions (\ref{eq:jump0}), (\ref{eq:jump1}) and
(\ref{far field}) with the stress jump ${\bf f}$ given by the constitutive equation (\ref{stressconstitutiveform}), together with a high-order Cahn-Hilliard type equation (\ref{eq:phase}) with the surface flux given by the constitutive equation (\ref{flux constitutive form}).

\subsection{Numerical Scheme}
Our numerical scheme is based on the work of \cite{Jinsun} with some minor improvements using ideas from \cite{marple2015fast}. We provide a brief summary of the method here. The governing equations for both the membrane phase evolution (\ref{eq:phase}) and the position evolution (\ref{interface equation}) are numerically stiff as they  contain  high-order spatial derivative terms. Consequently, explicit time-marching schemes tend to be prohibitively expensive requiring extremely small time-step sizes. To overcome the stiffness, our time-marching scheme uses the {\em small-scale decomposition approach} of \cite{HouT1997}, see \cite{Jinsun} for a detailed description of this method applied to vesicle flows. 

We employ the boundary integral method to solve the equations (\ref{eq:Stokes}), wherein, the velocity field at any point ${\bf x}$ is represented using a boundary integral as  
\begin{equation}
\label{eq:u1}
{\bf u} ({\bf x}) = {\bf u}_\infty({\bf x}) + \frac{1}{4\pi\eta}\int_{\bnd} {\bf G} ({\bf x}  - {\bf y}) \,
{{\bf f}}({\bf y}) \, d\gamma({\bf y}),
\end{equation}
where the free-space Green's function for the Stokes equation is given by \cite{Pozrikidis1992}
\begin{equation} {\bf G} ({\bf r}) = - \log \rho \, {\bf I} \,
+ \, \frac{{\bf r} \otimes {\bf r}}{\rho^2}, \quad \rho = |\!| {\bf r} |\!|_2. \end{equation}

By construction, the representation (\ref{eq:u1}) satisfies the Stokes equations and the boundary conditions  (\ref{eq:jump0}), (\ref{eq:jump1}) and (\ref{far field}) \cite{Pozrikidis1992}. Taking the limit as ${\bf x} \rightarrow \gamma$ and then applying the kinematic boundary condition (\ref{interface equation}) gives us a integro-differential equation for the membrane evolution. Following \cite{Jinsun,Hou}, we apply a second-order accurate linear propagator method to evolve the tangent angles of the material points on the membrane and subsequently a second-order Adams-Bashforth method to retrieve the corresponding positions.
 
\section{Results}
We investigate the influence of the elastic inhomogeneity, relative ratio of the surface lipid phase concentration and the external shear rate on the two-dimensional vesicle dynamics in simple shear flows. Another important parameter we vary is the {\em excess length} of the vesicle, defined as $\Delta = L/R - 2\pi$, where $L$ is the perimeter of the vesicle and $R$ is the radius of a circle that has the same enclosed area i.e., $R = \sqrt{A/\pi}$ given the vesicle area is $A$. In our numerical experiments, the initial vesicle shape is always an ellipse with a fixed perimeter, $L=2.6442$ (corresponding to that of a $\frac{1}{2}:\frac{1}{3}$ ellipse). The initial phase distribution $\psi$ is assumed to be a mixture of both phases:
  $\psi(\alpha,0)=\bar{\psi}+\delta(3\cos \,2\pi\alpha + 0.5\cos 6\pi\alpha + 0.5\cos(8\pi\alpha))$,
where  $\bar{\psi}$ is the average concentration varying between $0.25$ and $0.75$ accounting for different ratios of the surface phases,  $\delta$ is a small perturbation parameter, and $\alpha\in[0,1]$ parameterizes $\gamma$.  
Unless otherwise specified, we set the default values of some of the parameter as follows: 
$a_0 = 100, \quad \epsilon=0.04, \quad dt=10^{-5},  \quad \delta = 0.05$. 
The bending stiffness is a linear combination of the hard phase $B_1$ (denoted by blue color) and the soft phase  $B_2$ (denoted by red color) i.e.,  $B(\psi) = (1- \psi) B_1+\psi B_2$. We set $B_1 = 1$ and $B_2 = 0.1$ for all our tests. 
\subsection{Nearly circular vesicle in shear flow}
\label{nearcircular}

\begin{figure}[!htb]
 \begin{center}
\includegraphics[  width=2.5in]{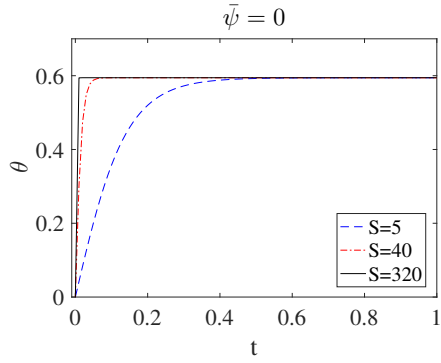}
\end{center}
\caption{Evolution of the inclination angles of a vesicle with $\Delta = 0.194$ and $\bar{\psi}=0$ under different shear rates.  We observe that regardless of the shear rate, they attain the same value over time, as was established both experimentally and numerically by previous studies.
\label{Oinclineangle0}}
\end {figure}

First we investigate the dynamics of a nearly circular vesicle, with $\Delta=0.194$, in shear flow by varying the shear rate $S$ and the average concentration $\bar{\psi}$. A well-known phenomenon in the case of a homogenous membrane is that when a vesicle with no viscosity contrast is subjected to shear flow, it undergoes a steady tank-treading motion. Moreover, the angle of inclination $\theta$ and the tank-treading frequency both are independent of the shear rate \cite{kraus1996fluid, Veerapaneni2009}. By setting $\bar{\psi} = 0$, we verify this result in Fig. \ref{Oinclineangle0}, where $\theta$ approaches the equilibrium value $\sim \! 0.6$  radians for $S \in[2,350]$.

\begin{figure}[!htb]
 \begin{center}
\includegraphics[ width=2.5 in]{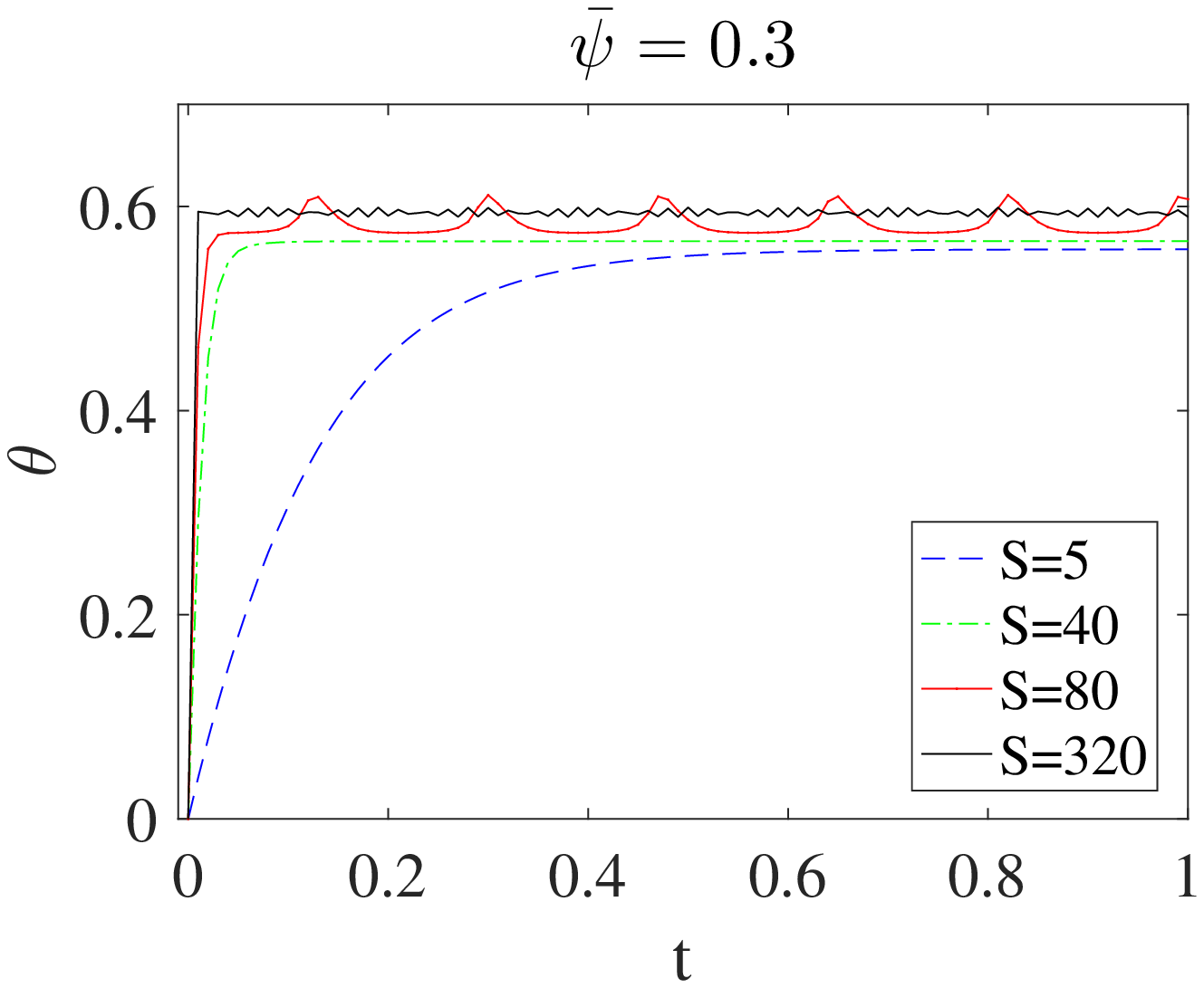}(a) 
\includegraphics[ width=2.5 in]{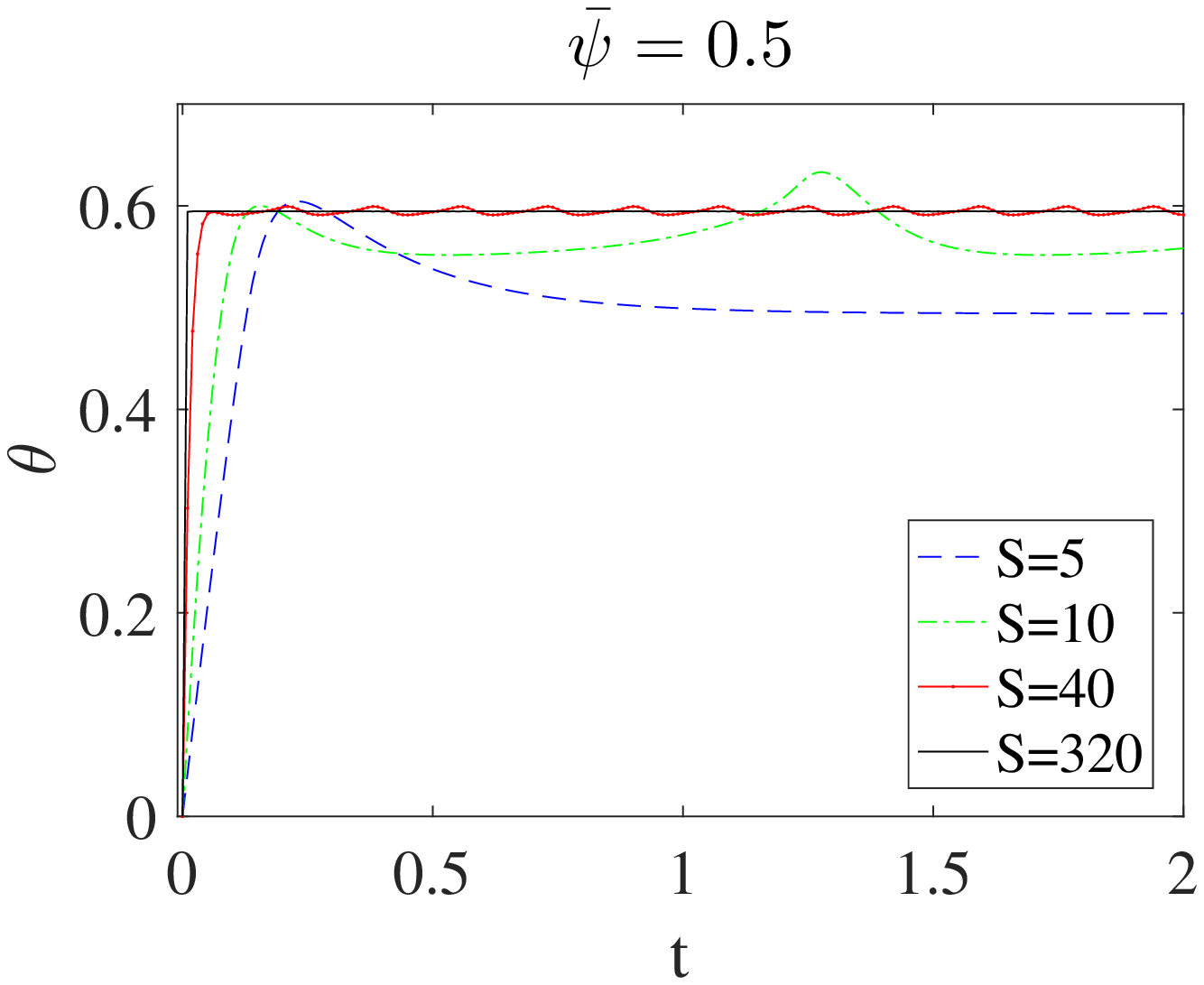}(b)
\includegraphics[ width=2.5 in]{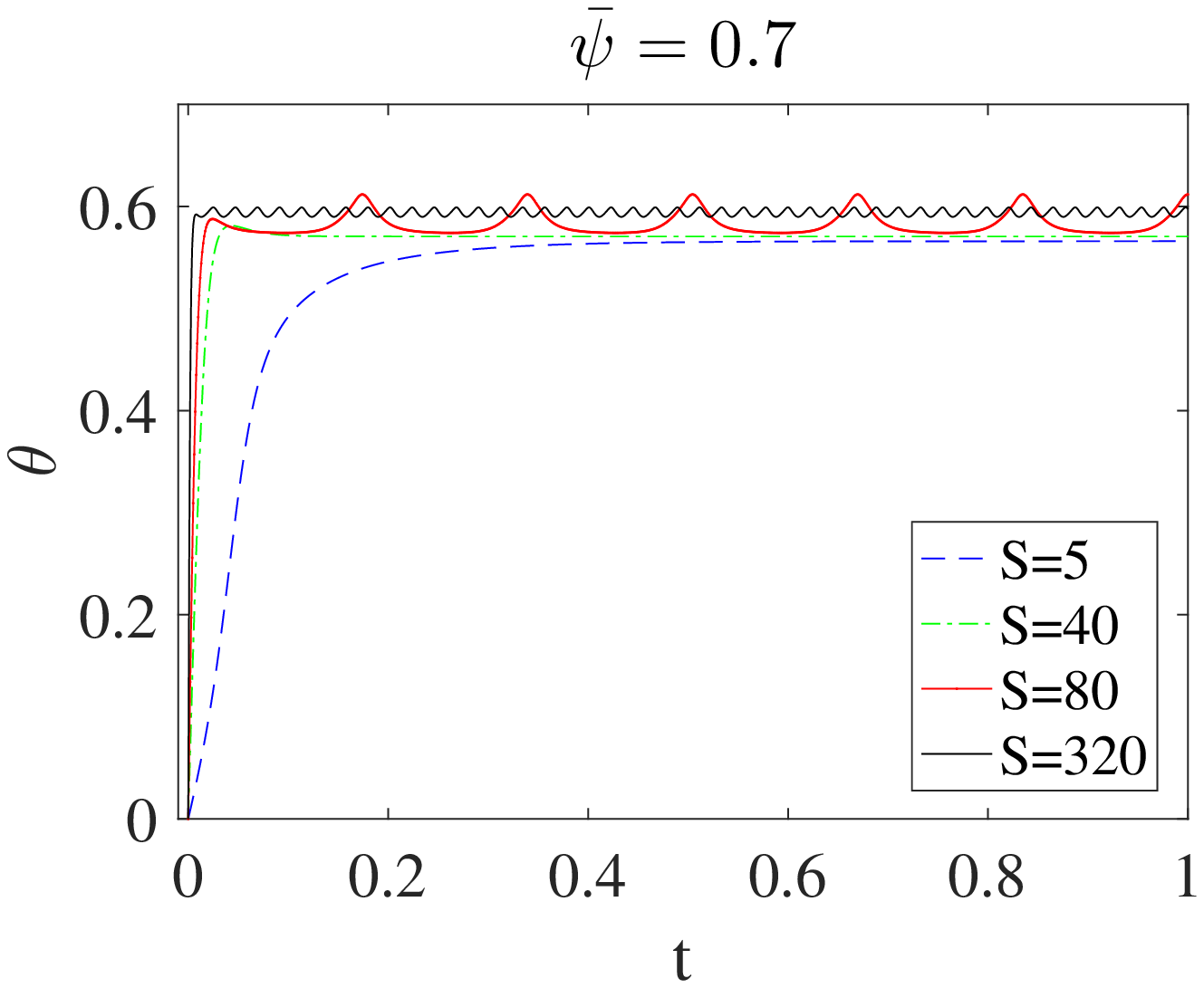}(c)
\end{center}
\caption{Evolution of the inclination angles of a vesicle with $\Delta = 0.194$ under different shear rates. The mean phase concentrations are given by (a)  $\bar{\psi}=0.3$, (b) $\bar{\psi}=0.5$, (c)  $\bar{\psi}=0.7$. 
\label{Oinclineangle1}}
\end {figure}

\begin{figure*}[!htb]
 \begin{center}
\includegraphics[ trim =50mm 5mm 50mm 5mm, width=5.2 in]{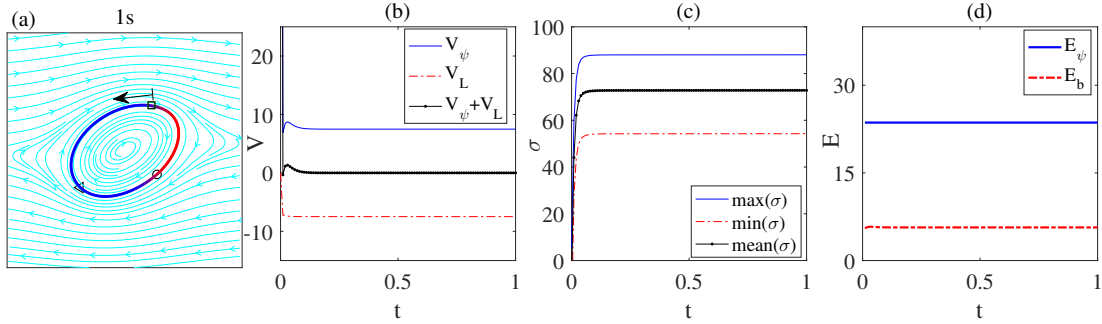} 
\end{center}
\caption{Dynamics of a tank-treading vesicle in shear flow with $\Delta=0.194$,  $\bar{\psi}=0.3$, and $S=40$. (a) Plot of the phase distribution and streamlines at equilibrium. (b) Velocity of the phase and the reference point. Here, $V_{\psi}$ is the velocity of the phase with respect to the reference point, $V_{ L}$ is the velocity of the reference point, and $V_{ \psi }+V_L$ is the velocity of the phase with respect to the Cartesian framework. Evolution of the (c) surface tension and (d) phase and bending energies.
\label{tanktreading}}
\end {figure*}

On the other hand, for a two-phase membrane, we find that the inclination angle does not remain fixed and is affected by the shear rate as shown in Fig. \ref{Oinclineangle1}, (a), (b) and (c). When $0.25<\bar{\psi}<0.75$, the initial mixture of membrane phases separates early on in the evolution, resulting in two large regions of the $\psi \approx 1$ (the red soft phase) and $\psi\approx0$ (the blue hard phase) (see Fig. \ref{tanktreading}).

\begin{figure*}[ht]
 \begin{center}
\includegraphics[ trim =50mm 5mm 50mm 5mm, width=5.2 in]{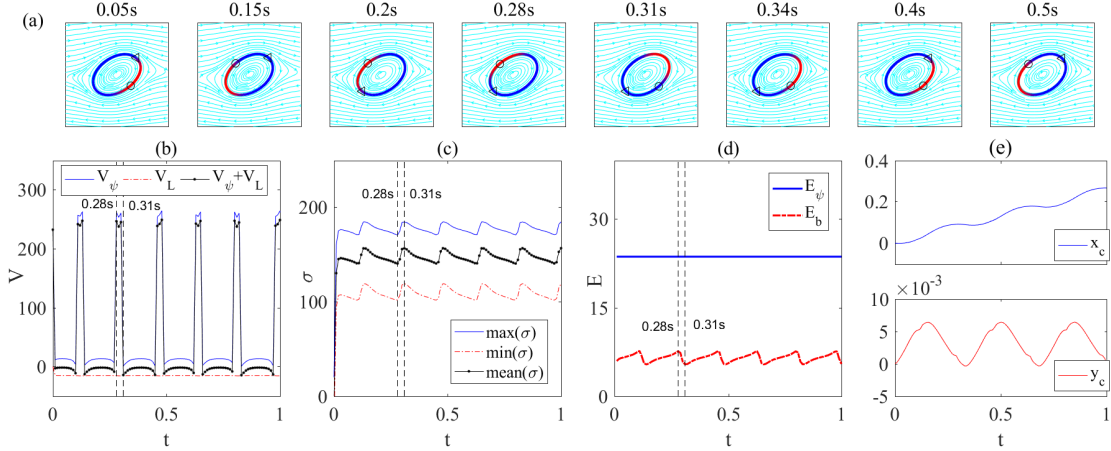} 
\end{center}
\caption{ (a) Snapshots from the simulation of a phase-treading vesicle suspended in shear flow with $\Delta=0.194$, $\bar{\psi}=0.3$, and $S=80$. Plots of evolution of the (b) velocity of the phase and the reference point, (c) surface tension, (d) membrane bending and phase energies and (e) centroid of the vesicle.
\label{phasetreading1}}
\end {figure*}

\begin{figure}[ht]
\begin{center}
\includegraphics[ width=2.5  in]{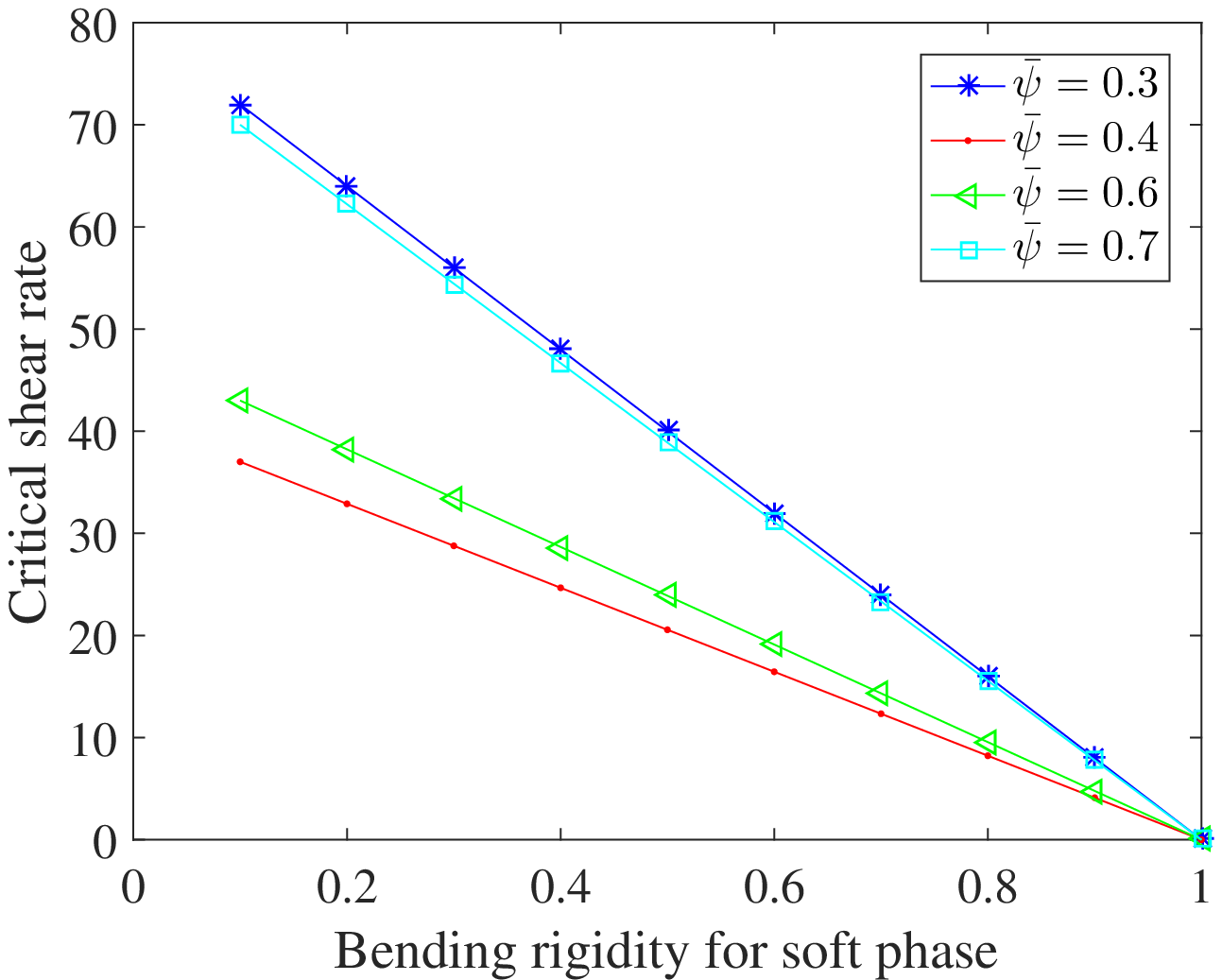}(a) 
\includegraphics[ width=2.5  in]{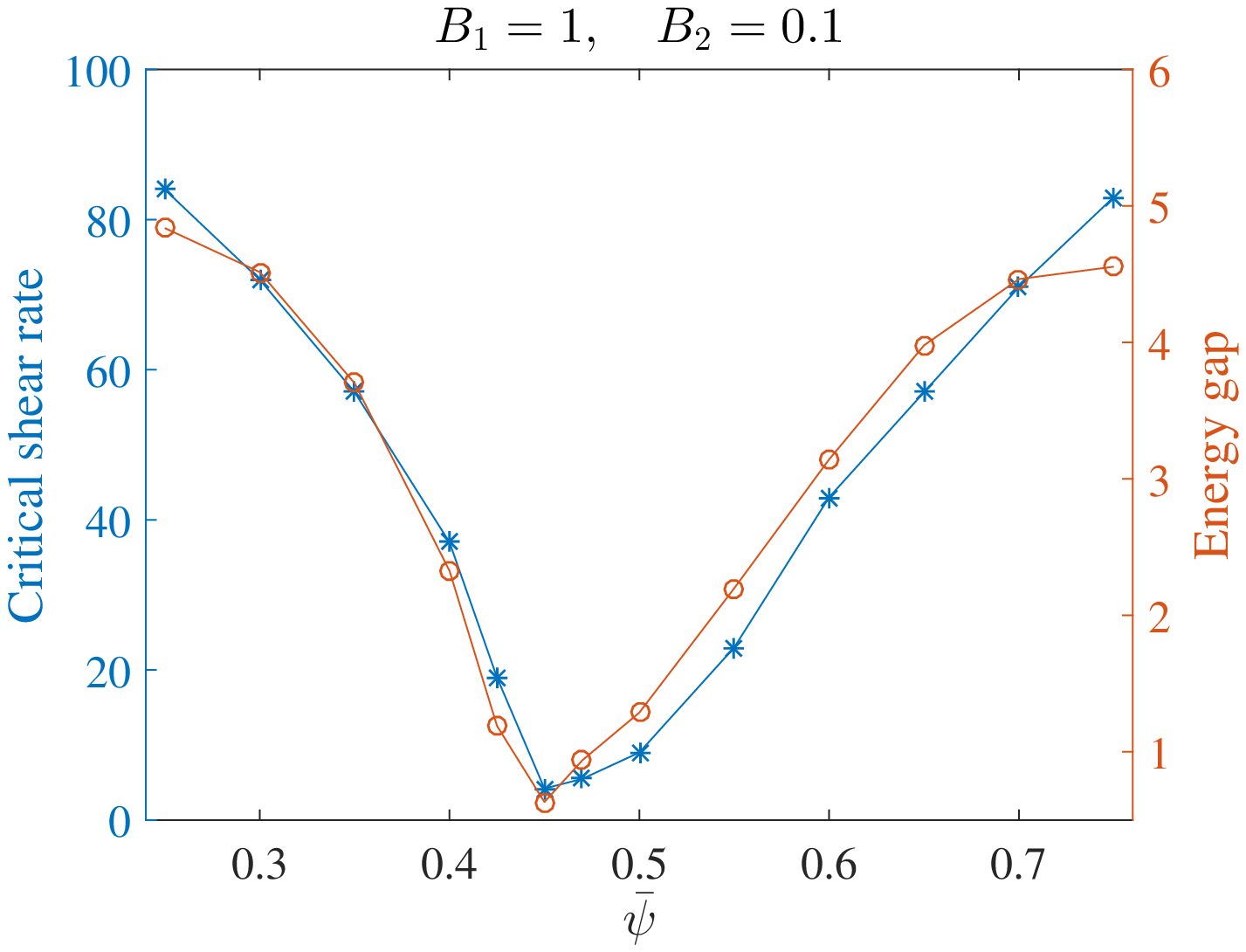}(b)
\includegraphics[ width=2.5  in]{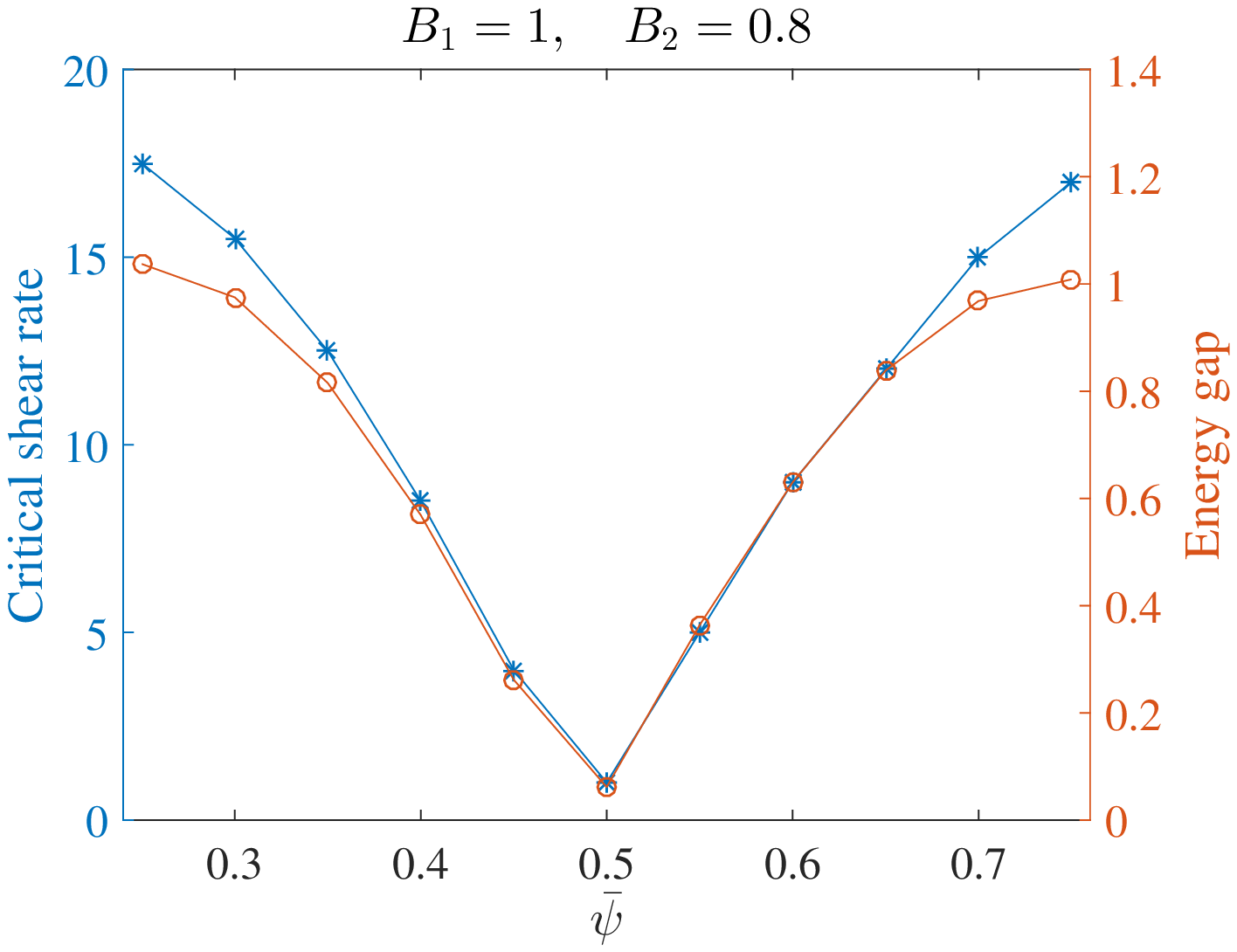}(c)  
\end{center}
\caption{Critical shear rates as a function of the bending stiffness $B_2$ of the soft phase (with $B_1$ fixed) in (a) and the average phase concentration in (b) and (c) (with different values for $B_2$). 
\label{criticalshearate}}
\end{figure}

Moreover, when the shear rate is small, the vesicle will tank-tread with the phase staying in place on the vesicle membrane (with respect to the fluid domain) and the variation of the inclination angle for different shear rates remains small. However, unlike the dynamics of a single phase vesicle, there exists a critical shear rate $S_C$, above which the phase will start to move and the vesicle shape and the inclination angle will undergo periodic oscillations accordingly. We shall call this kind of dynamics as ``{\em phase-treading}'',  to distinguish from the well-known tank-treading dynamics.   Note that the variance of the inclination angle over time will decrease as the shear rate increases, as shown in Fig. \ref{Oinclineangle1}. This is because  the shape of the vesicle  is determined more by the shape parameter $\Delta$ and less by the bending rigidity under high shear rate.

We demonstrate the tank-treading dynamics of a multicomponent vesicle in Fig. \ref{tanktreading}. Fig. \ref{tanktreading}(a) shows the fluid velocity (denoted by the streamlines) , the shape of the vesicle, and the phase distribution (denoted by color) at the equilibrium state. The reference point $\alpha=0$ is highlighted by a black square; $\alpha$ increases counter-clockwise (denoted by the arrow at the reference point). In Fig. \ref{tanktreading}(b), we plot the velocity of the reference point $V_{ L}$,  the velocity of the phase with respect to the reference point in Lagrangian frame $V_{\psi}$, and the velocity of the phase with respect to the fluid $V_{ \psi }+V_L$. For tank-treading dynamics $V_{ \psi }+V_L=0$ in the equilibrium state.  Fig. \ref{tanktreading}(c) shows the evolution of the maximum, minimum, and average values of the local surface tension. As can be observed, they remain fixed once the vesicle reaches the equilibrium state as is the case with the membrane energies plotted in Fig. \ref{tanktreading}(d).  Note that we use  a black circle and triangle to denote the position of the maximum and minimum value of the local surface tension respectively in Fig. \ref{tanktreading}(a).  Surface tension is largest when the curvature is the lowest and vice versa. 

We demonstrate the phase-treading dynamics of a multicomponent vesicle in Fig. \ref{phasetreading1}.  Fig. \ref{phasetreading1}(a) shows the morphological evolution of the vesicle, the corresponding phase distribution and the flow field.  While the phase boundaries separating the two phases move along with the membrane, the ambient fluid flow remains nearly the same.  In Fig. \ref{phasetreading1}(b), we plot $V_{ L}$,   $V_{\psi}$, and  $V_{ \psi }+V_L$. Notice that when the soft phase passes through high-curvature regions e.g., between $t=0.28s$ and $t=0.31s$, the phase boundaries move much faster. This is driven by bending energy dissipation, as shown in Fig. \ref{phasetreading1}(d). 
 In Fig. \ref{phasetreading1}(c) we plot the evolution of the surface tension. Note that between time $t=0.28$ and $t=0.31$,  the surface tension changes from minimum to maximum; but the phase energy remains the same, as shown in Fig. \ref{phasetreading1}(d). Fig. \ref{phasetreading1}(e) shows the position of the centroid of the vesicle. The vesicle moves off the origin, since the shape of the multi-component vesicle becomes asymmetric (unlike the tank-treading case) and the y-component oscillates periodically as the vesicle phase-treads.

\begin{table}
\begin{center}
\begin{tabular}{ c | c | c c c}
 \hline
 $\bar{\psi}$  &  $S_C$ & $ S = 80$   & $S = 160$ & $S = 320$\\
  \hline
0.3 & 72& 0.1729 & 0.0477 & 0.0221 \\
0.5  & 9 & 0.0863 & 0.0429 & 0.0216 \\
0.7  & 71 & 0.1652 & 0.0478 & 0.0221 \\
\hline
\end{tabular}
\caption{Periods of the oscillatory phase-treading motion of a vesicle in shear flow. }
\label{table1}
\end{center}
\end{table}

As can be noticed from Figure \ref{phasetreading1}, a phase-treading vesicle reaches a dynamic equilibrium, wherein, the membrane variables such as the tension undergo a periodic oscillation. In Table \ref{table1}, we measure their periods of oscillation for different values of $\bar{\psi}$ and $S$. We observe that, asymptotically, the periods vary linearly with $1/S$ as the shear rate is increased.  

\paragraph*{Analysis of the critical shear rate.} The transition from the tank-treading to phase-treading dynamics in the case of a nearly circular vesicle is strongly dictated by the difference in the bending moduli of the two phases. This can be understood from the evolution equation of the phase distribution,
\begin{equation}
{\psi _t} = \frac{a}{\varepsilon }{\left( {g' - {\varepsilon ^2}{\psi _{ss}}} \right)_{ss}} + B' \left( {H_s^2 + H{H_{ss}}} \right).
\label{psit}
\end{equation} 
Since the shape of the vesicle and $\psi$ do not vary much during tank-treading, all the terms in the right hand side of the above equation remain fixed, rendering ${\psi _t}$ to vary as $B'$, which equals the difference between the bending moduli of the two phases. We demonstrate this using numerical experiments in Fig. \ref{criticalshearate}(a) where we vary one of the bending moduli and see how $S_C$ varies for different values of $\bar{\psi}$. In all the cases, we observe that $S_C \sim B_1-B_2$ as predicted from our above analysis. 

We can also relate the critical shear rate to the membrane energy at equilibrium as follows. Recall that in the phase-treading regime, the bending energy of the membrane oscillates periodically unlike the tank-treading regime  (e.g., see Figs.\ref{tanktreading}(d) and \ref{phasetreading1}(d)). 
It turns out that the maximum variation of the bending energy within each period of oscillation is directly proportional to the critical shear rate. Let us define this bending energy gap as 
\begin{equation}
\Delta E_b=\max (E_b (t)) - \min (E_b(t)),
\end{equation}
for all $t$ in one period of oscillation; for example, in Fig. \ref{phasetreading1}(c), the maximum and the minimum occur at 0.31s and 0.28s respectively.  In Fig. \ref{criticalshearate}(b) and (c), we plot the critical shear rate and the bending energy gap as a function of $\bar{\psi}$. Clearly, the critical shear rate is positively related to the bending energy gap. In Fig. \ref{criticalshearate} (c), we plot the critical shear rate versus the average phase concentration, keeping $B_1=1$ and $B_2=0.8$. By doing this, we eliminate most of the non-linear part caused by periodic change of the vesicle shape. The critical shear rate and the energy gap is symmetric with respect to $\bar{\psi}=0.5$. Still, the reason for the discrepancy between the critical shear rate and bending energy gap is not clear.

\subsection{Dynamics of an elongated multicomponent vesicle}

\begin{figure}[t!]
 \begin{center}
 \includegraphics[trim= 10  0  10   0,  width=2.4 in]{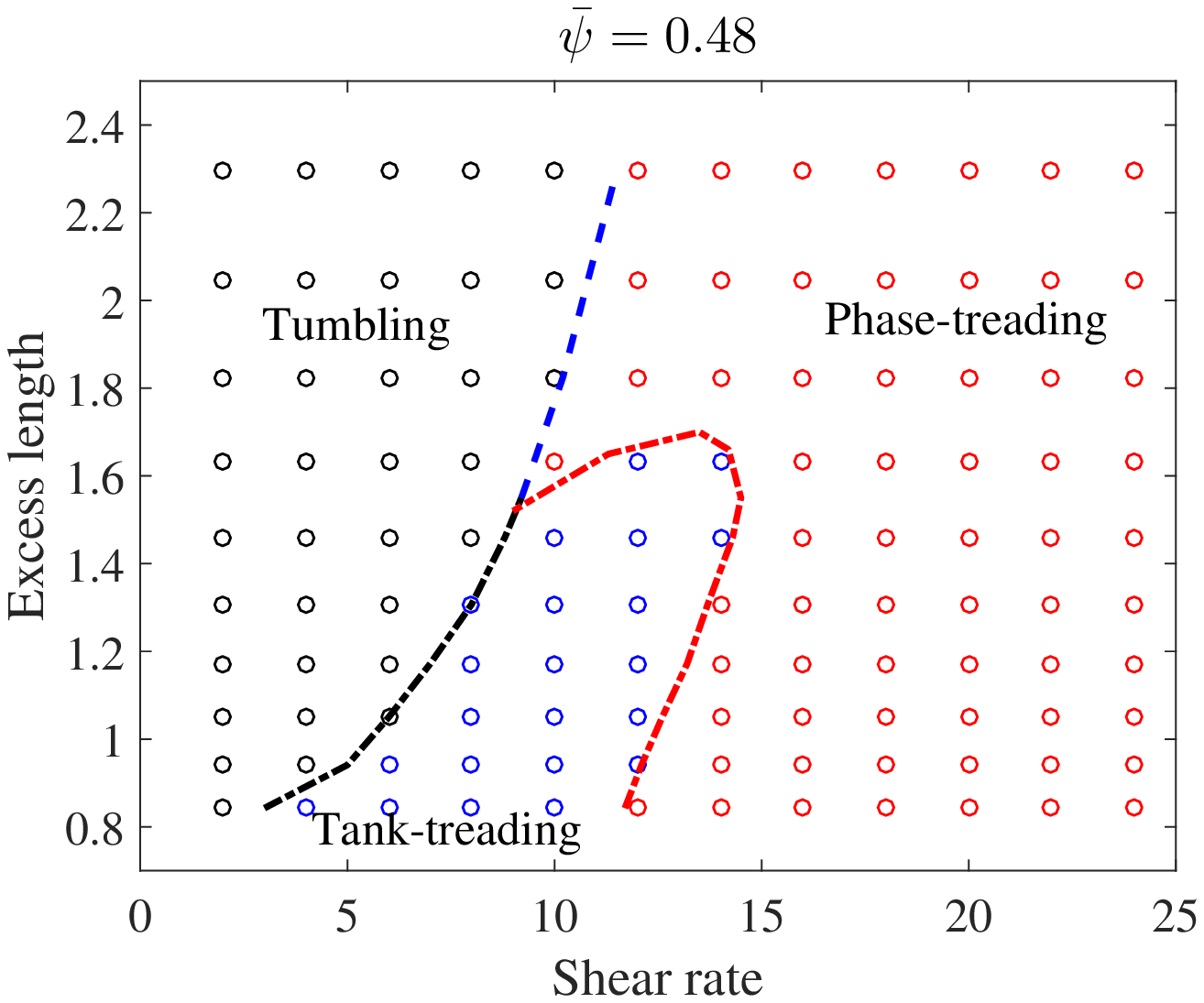} (a)
\includegraphics[trim= 10  0  10   0,  width=2.4 in]{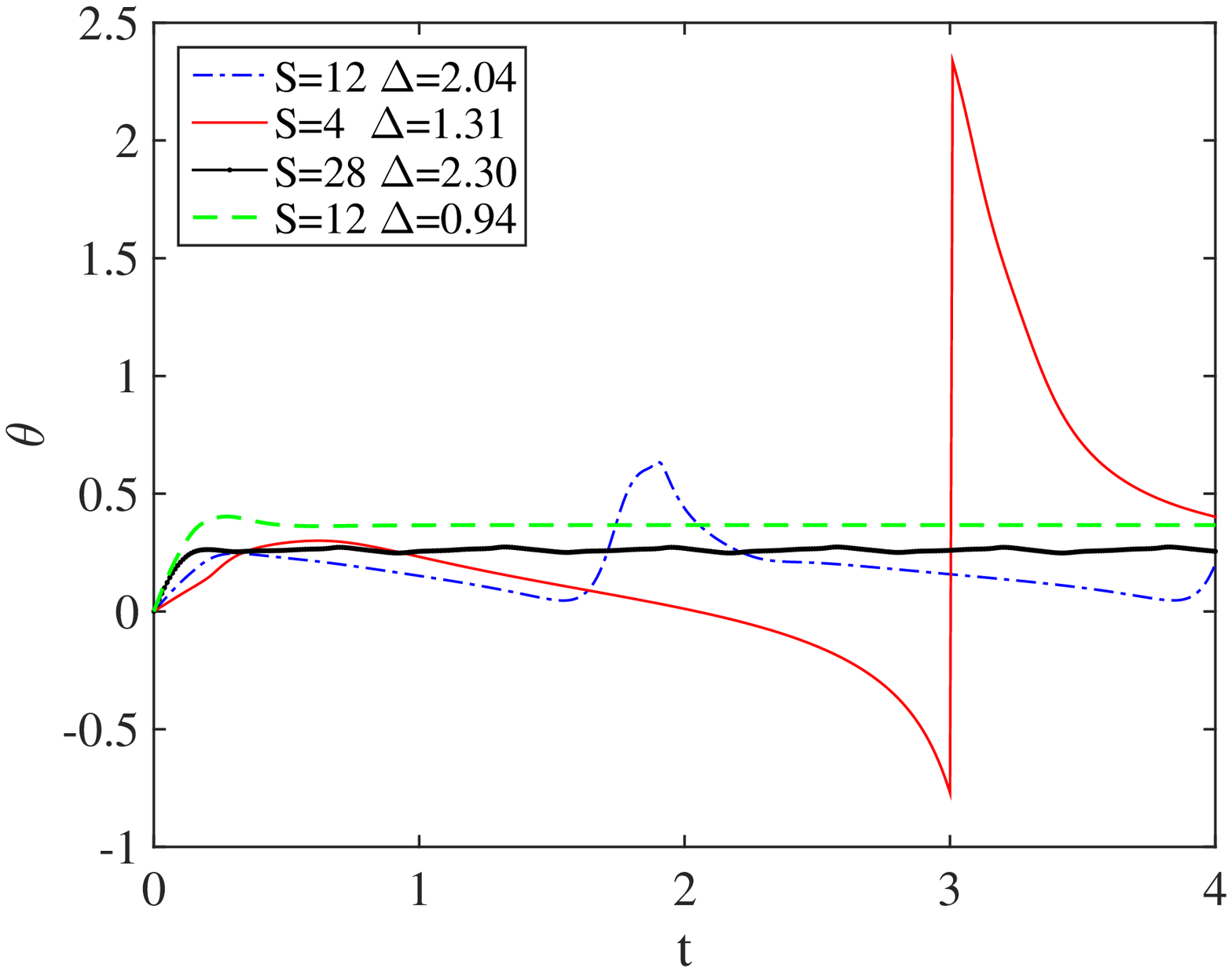}(b)
\end{center}
\caption{(a) Vesicle behavior as a function of the shear rate and excess length for $\bar{\psi}=0.48$. (b) Evolution of the inclination angle for four different cases. 
\label{phasediagram}}
\end{figure}
 
\begin{figure*}[ht]
\includegraphics[ trim= 50  0  50   0,  width=7.2 in]{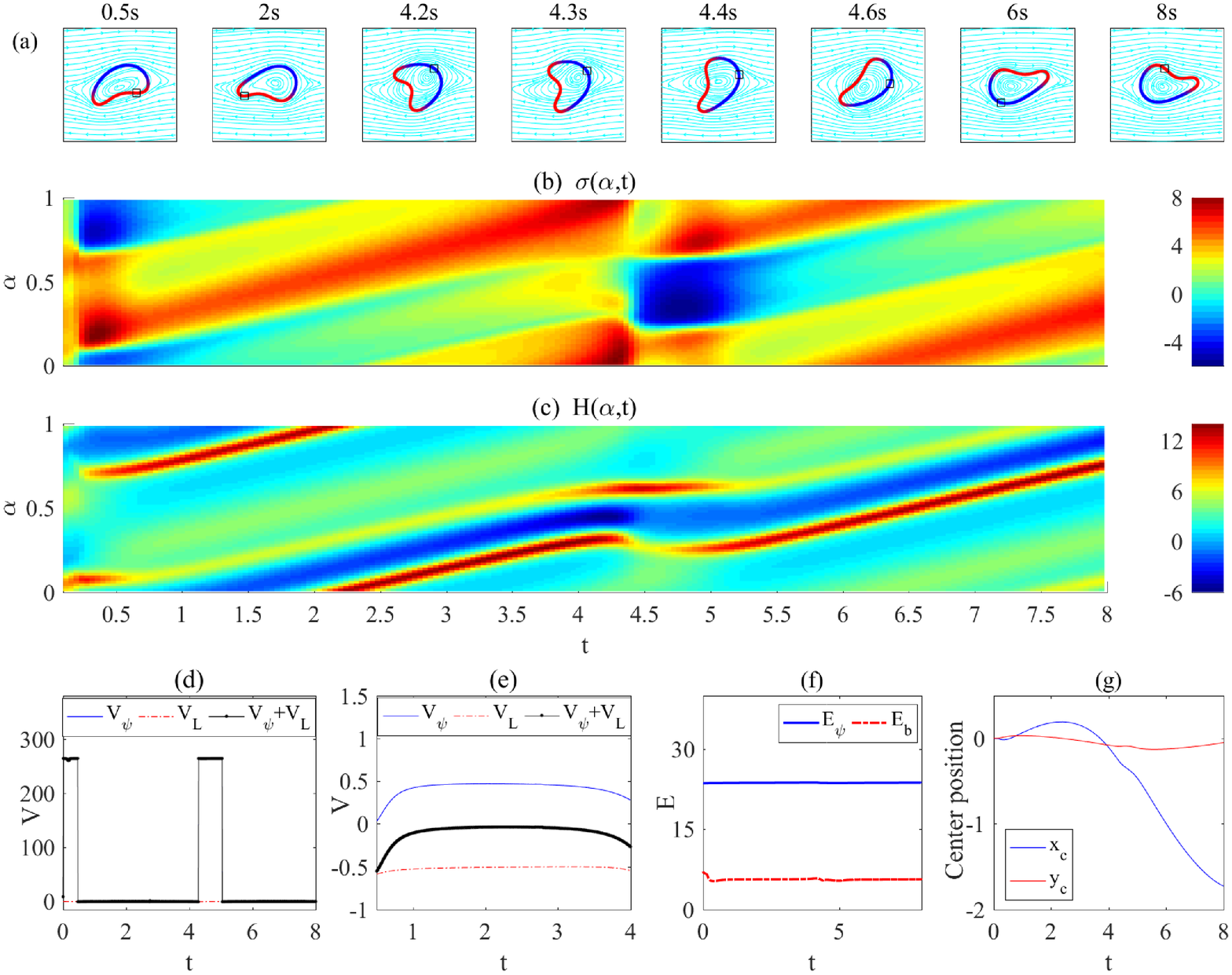} 
\caption{(a) Snapshots from the simulation of an elongated vesicle suspended in shear flow with $S=4$, $\Delta=0.94$ and $\bar{\psi}=0.48$. Evolution of the (b) surface tension, (c) curvature, (d) phase velocity, (e) phase velocity, (f) bending and phase energies, [g] centroid of the vesicle. 
\label{dynamicstu}}
\end{figure*}

\begin{figure*}[ht]
\includegraphics[ trim= 50  0  50   0,  width=7.2 in]{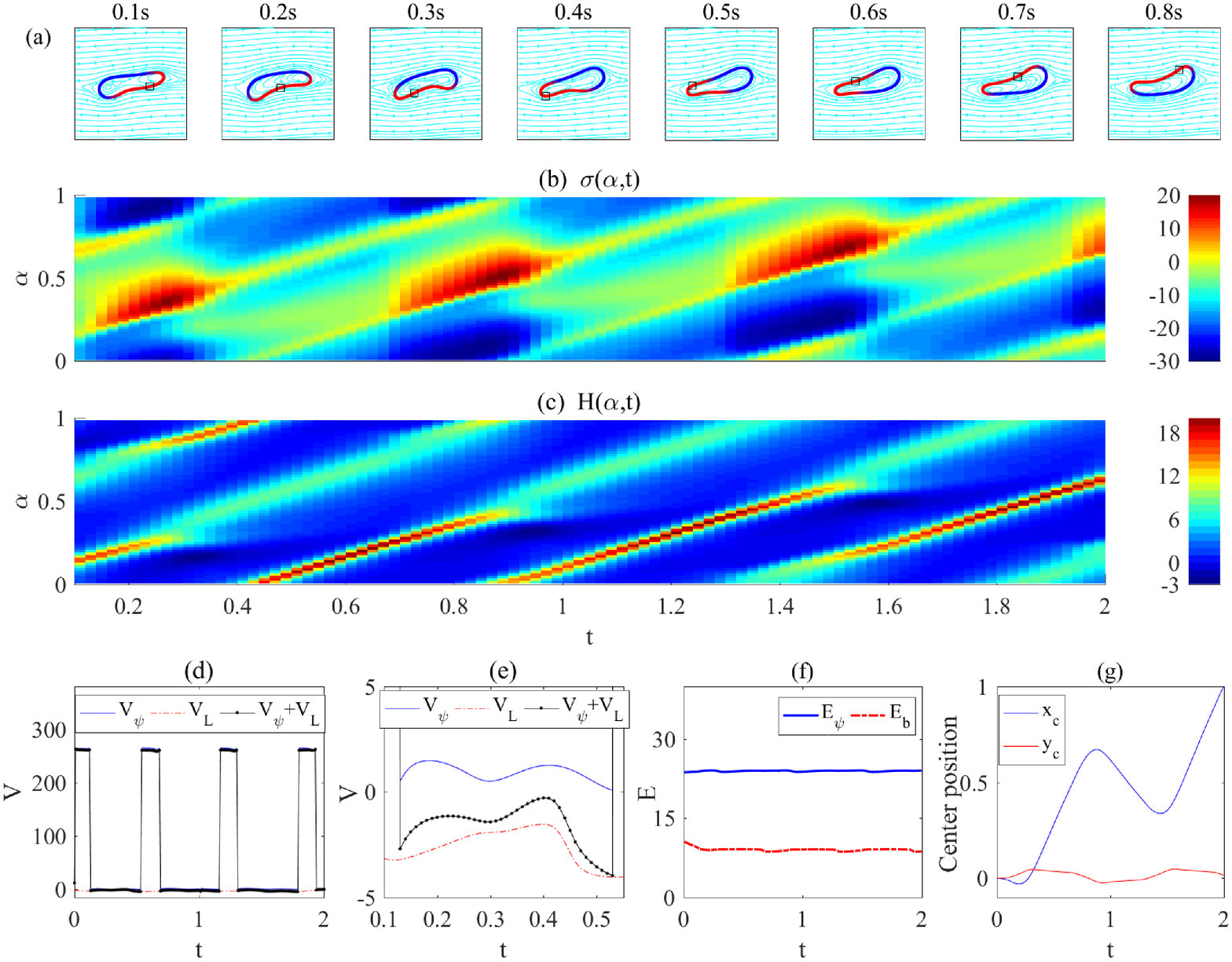} 
\caption{ 
(a) Snapshots from the simulation of an elongated vesicle suspended in shear flow with $S=28$, $\Delta=2.3$ and $\bar{\psi}=0.48$. Evolution of the (b) surface tension, (c) curvature, (d) phase velocity, (e) phase velocity, (f) bending and phase energies, [g] centroid of the vesicle. 
\label{dynamicspt}}
\end{figure*}

\begin{figure}[t!]
(a)\includegraphics[  width=2.6 in]{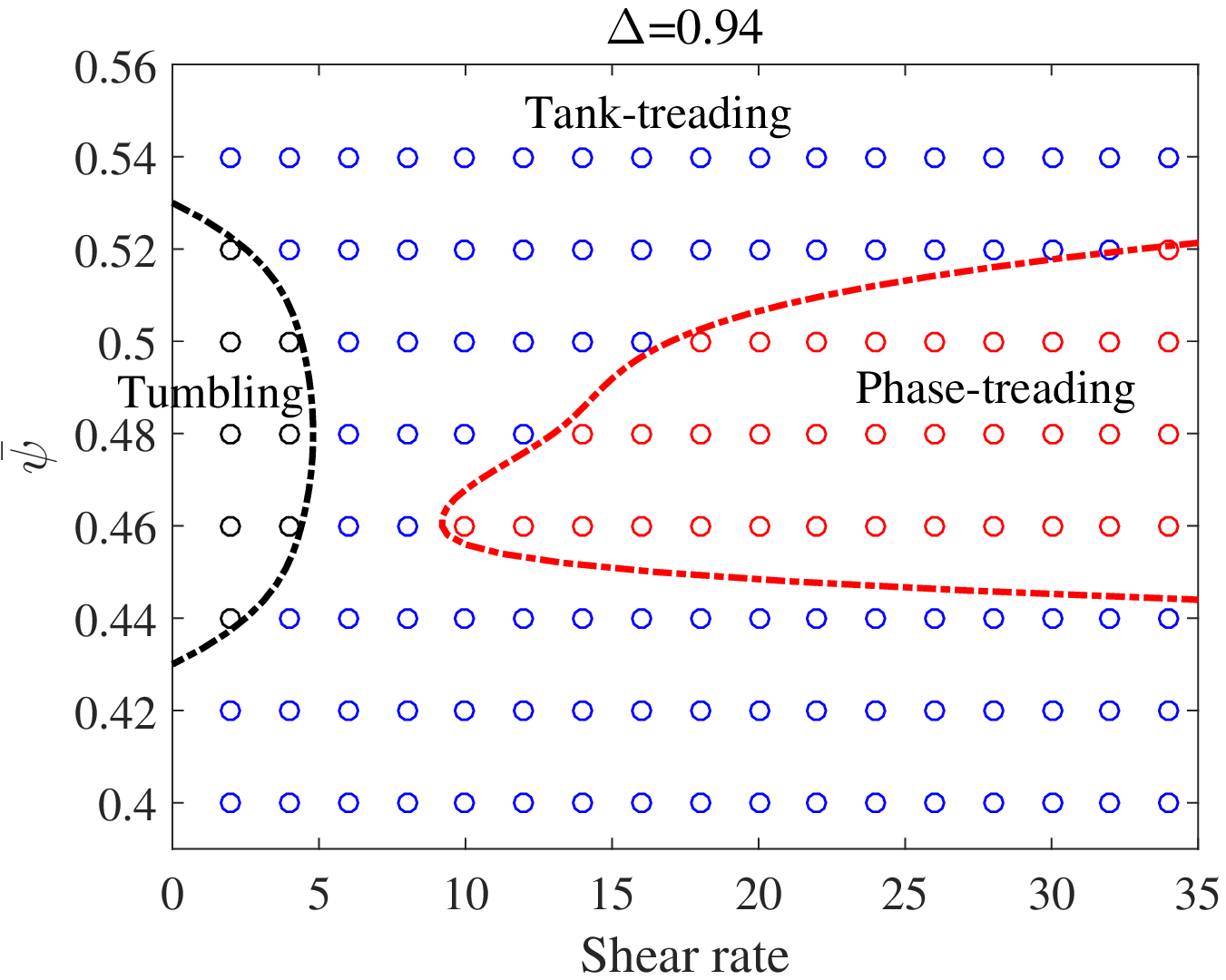}\\
\includegraphics[trim= 20  0  20   0,   width=3.2 in]{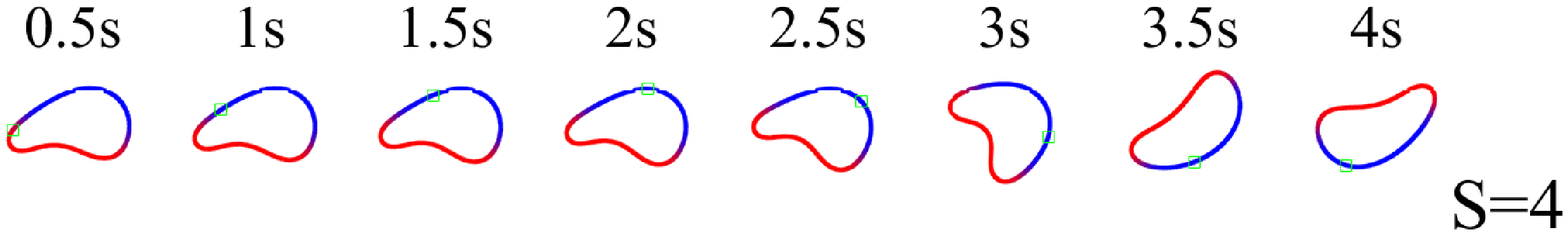}\\
\includegraphics[trim= 20  0  20   0,     width=3.2 in]{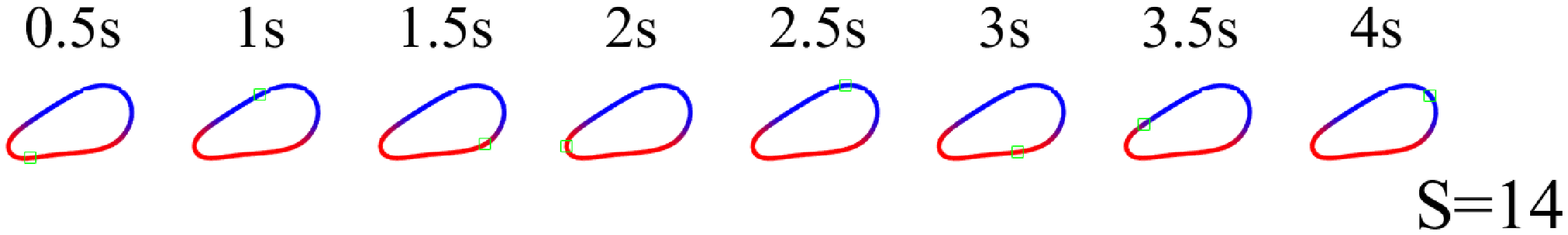}\\
\includegraphics[ trim= 20  0  20   0,    width=3.2 in]{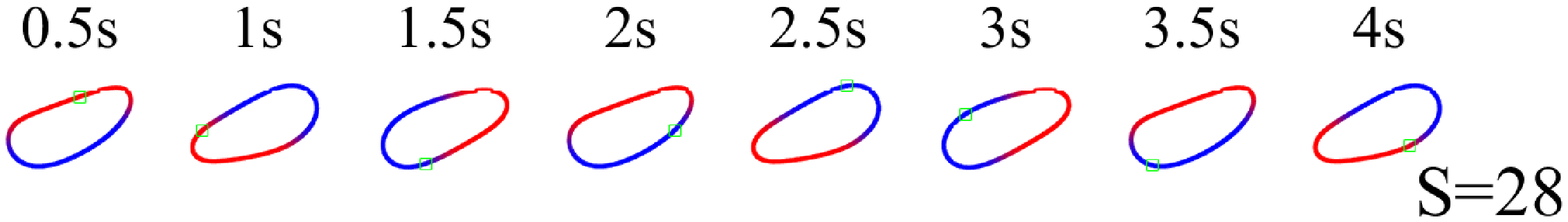}\\
(b)\includegraphics[    width=2.6 in]{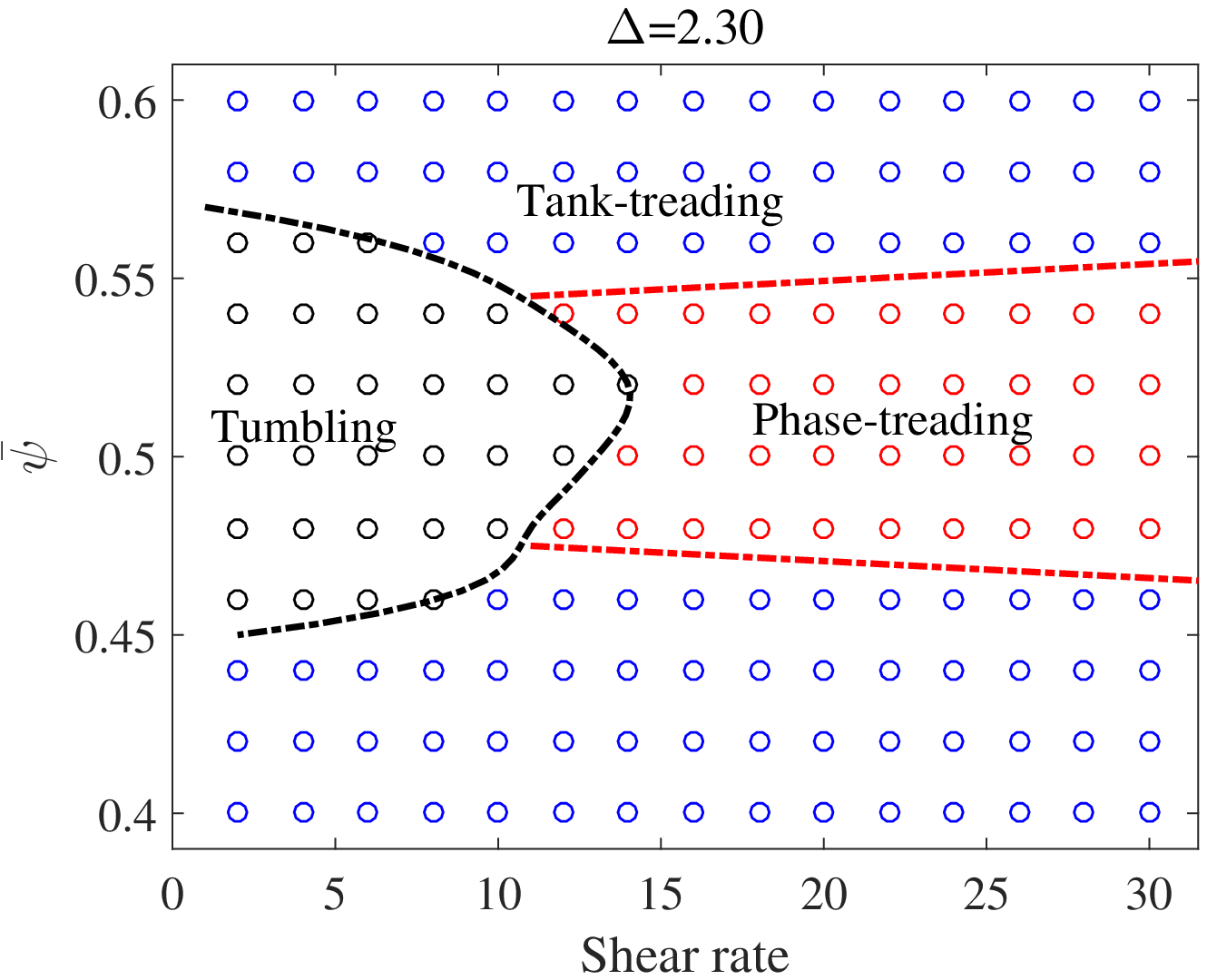}\\
\includegraphics[trim= 20  0  20   0,     width=3.2 in]{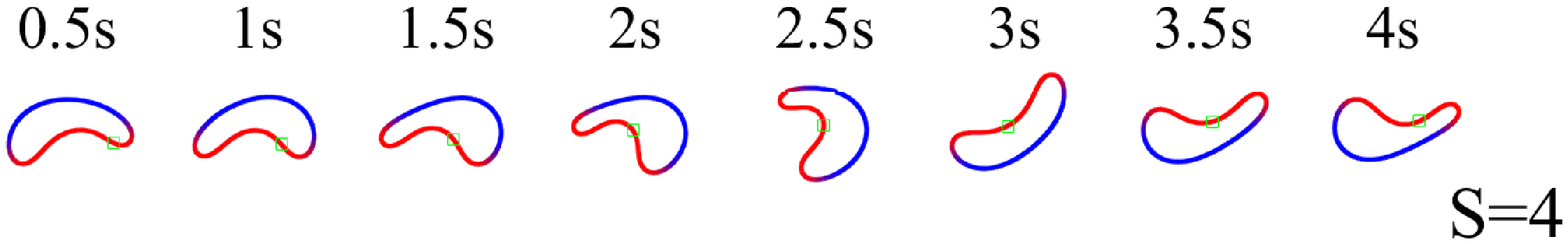}\\
\includegraphics[ trim= 20  0  20   0,    width=3.2 in]{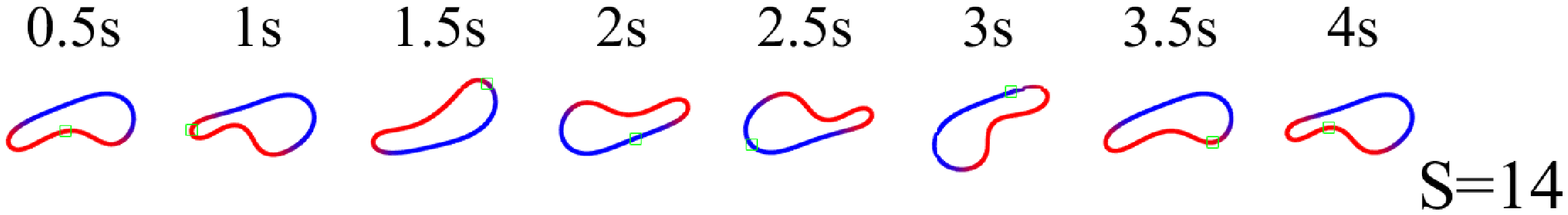}\\
\includegraphics[ trim= 20  0  20   0,    width=3.2 in]{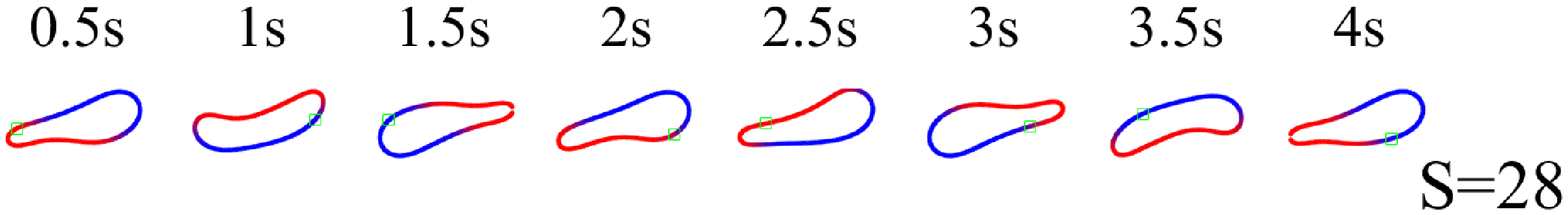}\\
\caption{
Phase diagram of vesicle dynamics as a function of shear rate and average phase concentration with (a) $\Delta=0.94$ and (b) $\Delta=2.30$.  Sequences of vesicle shapes when $\bar\psi=0.5$ under shear rate $S=4$, $14$, and $28$ for (a) $\Delta=0.94$ and (b) $2.30$.
\label{phasediagram2}}
\end{figure}

Next we investigate the dynamics of a more elongated vesicle in shear flow, by changing the shear rate $S$, the average concentration of the red soft phase $\bar{\psi}$, and the shape parameter $\Delta$.

In Fig. \ref{phasediagram} (a), we plot the phase diagram for the dynamics of multicomponent vesicle in shear flow, given fixed average phase concentration $\bar{\psi}=0.48$ and total arclength of the vesicle $L=2.6442$.  Unlike the nearly circular vesicle, the dynamics are now characterized by three regimes: (I) Tumbling: when the shear rate is small $S<4$ and the excess length $\Delta>0.8$; (II) Phase-treading: when the shear rate $S>15$ and the excess length $\Delta<3$; (III) Tank-treading: when $\Delta<1.6$. Note that when $\Delta>1.6$, there will be no tank-treading regime, and there is no sharp transition between the tumbling region and phase-treading region. The criterion we use to separate the tumbling from the phase treading region is that the minimum of the inclination angle can be negative, i.e. when $\min(\theta(t\rightarrow\infty))<0$  (see Fig. \ref{phasediagram} (b)), the vesicle is tumbling. The non-sharp condition means the $\min(\theta(t\rightarrow\infty))$  changes continuously with $S$.

In Fig. \ref{dynamicstu}, we plot the tumbling dynamics for a vesicle with excess length $\Delta=0.94$ when $S=4$.  It is well-known that a sufficiently  elongated vesicle of single component will tumble  when the shear rate is high .  To the best of knowledge, tumbling of a vesicle with relatively small excess length under low shear rate is new. The vesicle tumbles  because  it can bend inward on the soft phase region. In Fig. \ref{dynamicstu} (a), we plot the morphological evolution and the corresponding flow field. When $t=2s$ and $6s$, the vesicle  is along $x$ direction, that the nearby flow parallels the shape of the vesicle. For $t=4.2-4.6$s, the vesicle is along $y$ direction, and the flow around the vesicle membrane is rotating, i.e. the vesicle is tumbling. In Fig. \ref{dynamicstu} (b) and (c), we plot the evolution of the surface tension $\sigma(\alpha,t)$ and the mean curvature  $H(\alpha,t)$. For $t=1-4$s and $t=5-8$s, negative curvature, i.e. where the vesicle bend inward, the surface tension is small, even negative. The surface tension is always positive for the hard blue part, where the membrane bend outward. The surface tension is small(in absolute value) at the two tips, where the blue phase and read phase meet.   Note that there is a shape transition between $t=4.2-4.6$s. The phase moves quickly, as shown in Fig. \ref{dynamicstu}(d), and the bending energy pulses, as shown in  In Fig. \ref{dynamicstu} (f).  In Fig. \ref{dynamicstu}(e), we zoom in Fig. \ref{dynamicstu}(d) and show more clearly the relationship between $V_{ L}$,   $V_{\psi}$. For tumbling dynamics under lower shear rate,  $V_{ \psi }+V_L\approx 0$ most of the time, which is similar to tank-treading. The local surface tension  drops significantly (the blue region around $t=4.4$s and $\alpha=0.5$), which means that there is a compression of the vesicle.   Also  the vesicle moves off the center position as  the shape of the vesicle changes periodically, as shown in Fig. \ref{dynamicstu} [g].

  For smaller $\Delta$, i.e. $\Delta<1.6$,   as the shear rate increases, the vesicle will tank-tread. The vesicle will finally phase-tread if we keep increasing the shear rate. There will also be a critical shear rate $S_C$ under which the vesicle will tank-tread, as shown in Fig. \ref{phasediagram}, (a). However, for an elongated vesicle the $ S_C\sim B_1-B_2$ relation will not hold anymore since the high curvature $\max_{\alpha} (H(\alpha)) $ will change significantly for different $B_2$.

 In Fig. \ref{dynamicspt}, we plot the phase-treading dynamics for a more elongated vesicle with excess length $\Delta=2.30$ when $S=28$. The phase move along the vesicle periodically and the shape of the vesicle changes periodically. As shown in Fig. \ref{dynamicspt} (a), where we plot the morphological evolution and the corresponding flow field,  the flow field agrees with the vesicle shape. In Fig. \ref{dynamicspt} (b) and (c), we plot the evolution of the surface tension $\sigma(\alpha,t)$ and the mean curvature  $H(\alpha,t)$. Between $t=0.4-0.6$s, the red phase moves across the left tip, and the surface tension is small (in absolute value). Between $t=0.7-0.9$s, when the red is on the top and blue phase is at the bottom, the surface tension is large (in absolute value). 
The surface tension is negative at the top where the vesicle bend inward, and the surface tension is positive at the blue bottom where the vesicle bend outward a little. Moreover, the vesicle moves off the center position as  the shape of the vesicle changes periodically, as shown in Fig. \ref{dynamicspt} [g].


Next we investigate the influence of the average phase concentration.  In Fig. \ref{phasediagram2}, we construct the phase diagram for the dynamics of a multicomponent vesicle in shear flow, given fixed total arclength  $L=2.6442$ and excess length $\Delta=0.94$ for Fig. \ref{phasediagram2} (a), and $\Delta =2.30$ for Fig. \ref{phasediagram2} (b), while changing the average phase concentration and the shear rate.
 
In Fig. \ref{phasediagram2} (a), there are 3 regions: (I) tumbling: when the shear rate  is small $S<4$ and the average phase concentration is around $0.5$, the vesicle will tumble; (II) tank-treading: when shear rate increases, the vesicle will tank-tread; (III) phase-treading: as the shear rate increases, the phase will tread along the vesicle together with the reference point.
 
The vesicle is more unstable if the average phase concentration is around $0.5$. On the one hand, the bending energy gap $\Delta E_b$ is smaller that the critical shear rate between the tank-treading and phase-treading region is smaller. On the other hand, when the shear rate is small, the half soft/half hard phase distribution can lead to bean shaped vesicle that will tumble in shear flow, which is significantly different from single phase vesicle.
 
In Fig. \ref{phasediagram2} (b), there are also 3 regions: (I) tumbling: when the shear rate  is small $S<4$ and the average phase concentration is around $0.5$, the vesicle will tumble; (II) phase-treading: when shear rate increases the average phase concentration is around $0.5$, the phase will tread with the reference point while the transition between the tumbling and phase treading is not sharp; (III) tank-treading: if the average phase concentration is large $\bar\psi>0.58$ or small $\bar\psi<0.44$, the vesicle will tank-tread within a wide range of shear rate, i.e. when $\bar{\psi}=0.44$, the critical shear rate between tank-treading and phase treading is $S_C=120$,  because the vesicle is more elongated that the bending energy gap $\Delta E_b$ is large. 
 
\section{Conclusions and Discussion}
We investigated the dynamics of two-dimensional multicomponent vesicles in shear flows with matched viscosity of inner and outer fluids. We  focused our study on how the inhomogeneous bending, the excess length and the  applied  shear rate dictate the vesicle dynamics.  Unlike the homogeneous vesicle dynamics, here we found three dynamic patterns---tank-treading, phase-treading, and tumbling. The critical shear rate for transitioning to phase-treading is shown to be proportional to difference in bending moduli as well the bending energy gap. This fact can come in handy, perhaps, in estimating the material properties of vesicle membranes by observing their dynamics in shear flows.    

For the two-phase vesicle considered in this work, the phases separate into two large regions, where the hard phase congregates at the low curvature region, and the soft phase congregates at high curvature region. This asymmetry in the shape often leads to a cross-streamline migration of the vesicle and a net preferential direction of motion even in linear shear flow. Elongated vesicles undergo tumbling when the shear rate is small if the average phase concentration is around 1/2. Moreover, the excess length required for tumbling is significantly smaller than the value for a single phase case. Similar to the phase-treading regime, the asymmetry in the shape of a vesicle can lead to a preferential migration since the centroid of the vesicle undergoes periodic oscillations. As the shear rate is increased, an elongated vesicle will transition from tumbling to tank-treading to phase-treading dynamics. 
Nevertheless, as the excess length is increased, the tank-treading region shrinks and finally disappears. Without the tank-treading regime, the transition from tumbling to phase-treading becomes blurry, since our criterion to distinguish them is whether the inclination angle can be negative.

The results presented here add to the increasing body of evidence that shows the importance of material composition on the dynamics of biomembranes at small scales. In general, we find that the most important effect of compound phase domains is to provide a symmetry-breaking source for the vesicle interface, and the effects of such an asymmetry on the dynamics are profound. When the excess length is small, the results are expected because the shape of the vesicle is not symmetric due to inhomogeneous bending and two phases will finally move with the nodes under strong shear flow. When the excess length is large, the influence of the inhomogeneous bending is more significant since the vesicle with evenly distributed phases can bend inward and tumble under weak flow.  

Several assumptions were made in this work. The bending modulus for the composite phase was assumed to be a linear combination of the bending moduli for each one; more complex models can easily incorporated in our simulation framework. 
We restricted our attention to parameter regimes where the phase energy is significantly larger than the bending energy, so that the phase boundaries separating the two phases are sharp and the phase distribution is relatively stable. The interior and exterior fluid viscosities were assumed to be the same; the inhomogeneous bending can lead to interesting interplay of the tumbling dynamics known to arise from viscosity contrast.  In addition, we are currently working on extending this work to three-dimensions and multi-particle interactions in confined flows.

\section{Acknowledgements}
K. L.,  S. L. and J. L. acknowledge the support from the National Science Foundation, Division of Mathematical Sciences (NSF-DMS) grants DMS-R6376(J. L.), DMS-0915128(J. L.), DMS-0914923(S. L.) and DMS-1217277(S. L.). S. L. is also partially supported by grant ECCS-1307625. S.V. thanks the National Science Foundation for partial support from a NSF Career grant DMS-1454010. K. L. and J. L. also thank partial support from the National Institutes of Health through grant P50GM76516 for a Center of Excellence in Systems Biology at the University of California, Irvine. Some computations in this work were performed on computers acquired using NSF grant (SCREMS) DMS-0923111. 

\bibliographystyle{plain}
\bibliography{SoftMatterpaper}

\end{document}